\documentclass[a4paper,12pt]{article}
\usepackage {amsmath}
\usepackage {amsxtra}
\usepackage {amsbsy}
\usepackage {amscd}
\usepackage {latexsym}
\usepackage {graphicx}
\usepackage {graphics}
\usepackage {a4wide}
\usepackage {amssymb}
\usepackage {amsfonts}
\usepackage {float}
\usepackage {indentfirst}

\begin{document}


\title{\bf Beta decay in external field and neutrino mass}

\author{\bf O. F. Dorofeev \;\;\; and\;\;\;
\bf A. E. Lobanov\thanks{\emph{E-mail:
lobanov@th466.phys.msu.su}}}
\date{}
\maketitle

\vspace{-7mm}
\begin{center}\emph{Moscow State University, Department of Theoretical
Physics, 119992 Moscow, Russia}%
\end{center}

\begin{abstract}
The results of the investigation of electromagnetic field effects
on the process of beta decay are used for analyzing experimental
data on direct neutrino mass search.
\end{abstract}

The investigation of the effect of  electromagnetic fields on the
process of beta decay was started  long ago. In particular, the
effect of plane-wave fields on this process was discussed in
\cite{L7,L10,L18}. The obtained results indicate that beta
spectrum is strongly affected by electromagnetic radiation. On the
other hand spectral distribution of $ \beta $-electrons is most
sensitive to neutrino mass $m _ {\nu} $. In papers
\cite{L7,L10,L18} the emphasis was made on very strong fields. In
our work we consider very weak fields and their effect on beta
spectrum.

Let us choose the monochromatic circularly polarized plane wave
with frequency $\omega$ and intensity $E$ as the model of external
field and suppose that $\;\omega\ll m_{\nu}.$  Then the total
probability of the allowed $\beta$-decay for the model with
massive Dirac neutrino is:
\begin{equation}\label{p2}
\frac{W}{
\tilde{W}}=\frac{\xi^{\,2}}{4}\left\{\int\limits_{{t}_{1}}^{
{t_{2}}}\! dt\!\!\int\limits_{y_{1}}^{\varepsilon_{0}-\mu}\!\!
dy\,{\Phi}(t,y)\,+\,\Theta(\xi_{0}-\xi)\!\int\limits_{{t}_{0}}^{
{t_{1}}} \!dt\!\int\limits_{y_{1}}^{y_{2}}
\!dy\,{\Phi}(t,y)\right\}.
\end{equation}
Here
\begin{equation}\label{p3}
\begin{array}{c}
\displaystyle {\Phi}(t,y)=
{(t+y)y(\varepsilon_{0}-y)\left[(\varepsilon_{0}-y)^{2}-
\mu^{2}\right]^{1
/\,2}}{\left[\xi^{2}+(y-t)^{2}\right]^{-3/2}}\,,\\[8pt]
\displaystyle  t_{0}=1,\quad \xi_{0}=\big[2(\varepsilon_{0}-\mu
-1)\big]^{1/2},\quad \beta=(1-1/t^{2})^{1/2},
\end{array}
\end{equation}
where
\begin{equation}\label{p1}
  \xi =eE/m\omega,\quad \mu =m_{\nu}/m,\quad
  t=p^{0}/m,\quad\varepsilon_{0}=(M_{i}-M_{f})/m,
\end{equation}
and $e,m,p^{0}$  are the charge, mass and total energy of
$\beta$-electron. The lower and upper limits of the integral
(\ref{p2}) are given by
\begin{equation}\label{p4}
\begin{array}{c}
{t}_{1,2}=(\varepsilon_{0}-\mu)(1+\xi^{2}/2)\mp
\xi(1+\xi^{2}/4)^{1/2}\left[
(\varepsilon_{0}-\mu)^{2}-1\right]^{1/2},\\[8pt]
{y}_{1,2}=t\big[1 + \xi^{2}/2\, \mp
\xi(1+\xi^{2}/4)^{1/2}\beta\,\big].
\end{array}
\end{equation}
For the neutron and approximately for tritium we have
$\tilde{W}=G^{2}_{F}m^{5}(1+3\alpha_{0}^{2})/2\pi^{2},$ where
$\alpha_{0}$ is the ratio of the axial and vector constants of
weak interaction, and $G_{F}$ is the Fermi constant.

\begin{figure}[H]
\begin{center}
\includegraphics[width=0.7\textwidth]{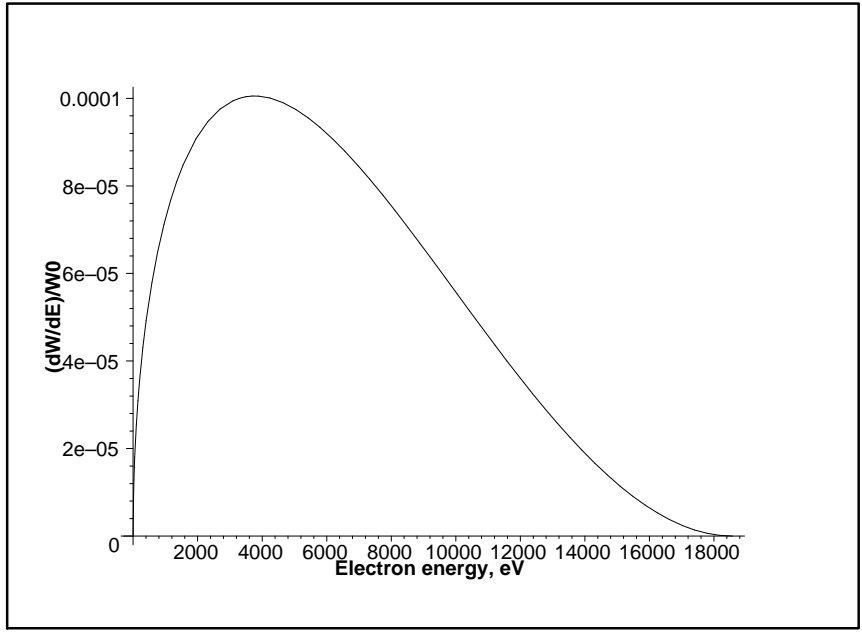}
\caption{{Tritium beta spectrum.}}\label{fermi1}
\end{center}
\end{figure}

\begin{figure}[H]
\begin{center}
\includegraphics[width=0.67\textwidth]{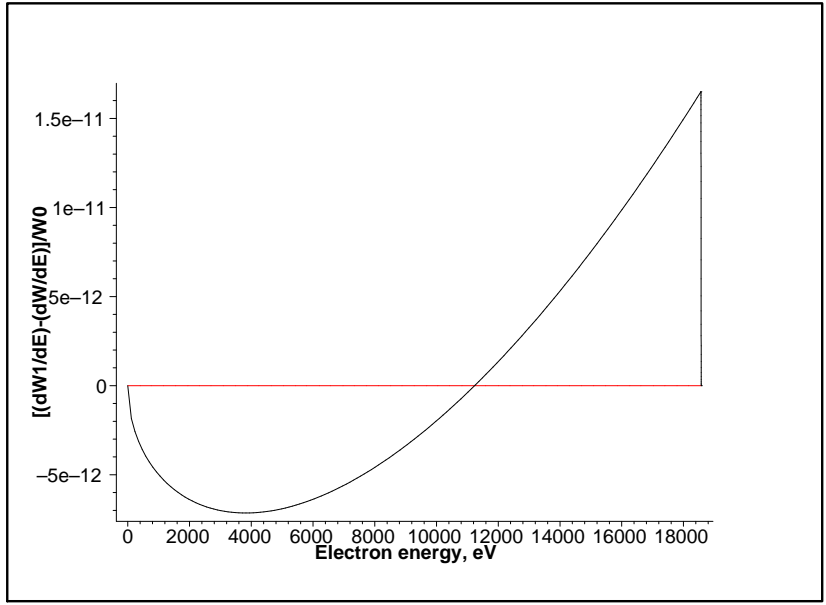}
\caption{{Tritium beta spectrum: $m_{\nu}=0;$ $\;W$ corresponds to
$\;\xi=0,$ $W_{1}$ corresponds to $\;
\xi=0.00005.$}}\label{fermi3}
\end{center}
\end{figure}

\begin{figure}[H]
\begin{center}
\includegraphics[width=0.67\textwidth]{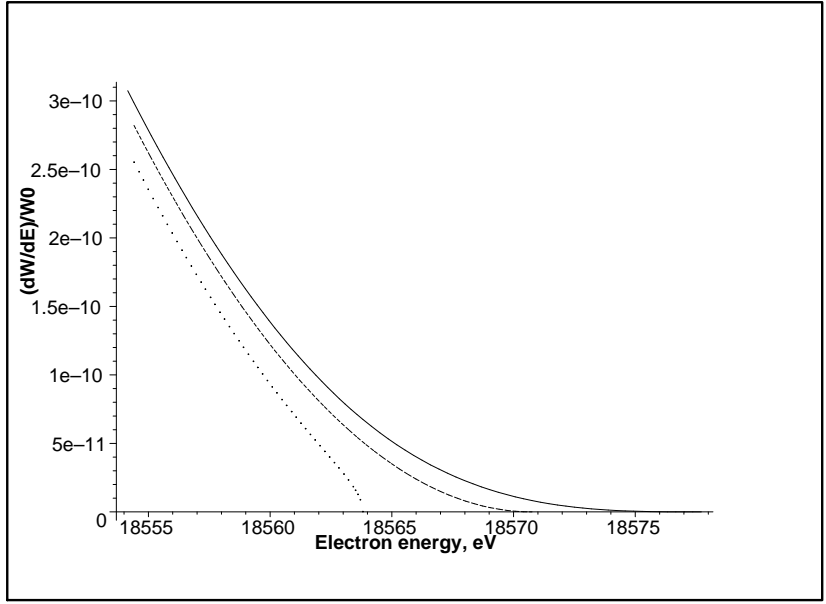}
\caption{Tritium beta spectrum near its end point: dashed line
corresponds to $m_{\nu} =0,$ $\xi=0; \; $ dotted line corresponds
to $m _ {\nu} =6.8 \, {\textmd {eV}}, \; \xi=0; \; $  solid line
corresponds to $m_{\nu}=0,$ $\xi=0.00005.$ }\label{fermi001}
\end{center}
\end{figure}

\begin{figure}[H]
\begin{center}
\includegraphics[width=0.67\textwidth]{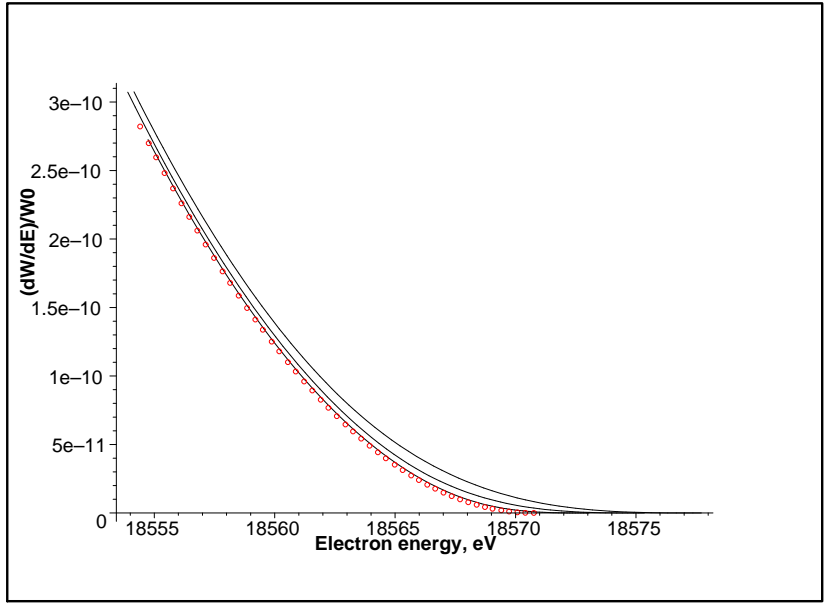}
\caption{Tritium beta spectrum near its end point: circles
correspond to $\xi=0,\;m_{\nu}=0;$ solid lines correspond to
$m_{\nu} = 0,\;\xi
=0.00005;\;0.000034;\;0.000017.$}\label{fermi07}
\end{center}
\end{figure}

\begin{figure}[H]
\begin{center}
\includegraphics[width=0.67\textwidth]{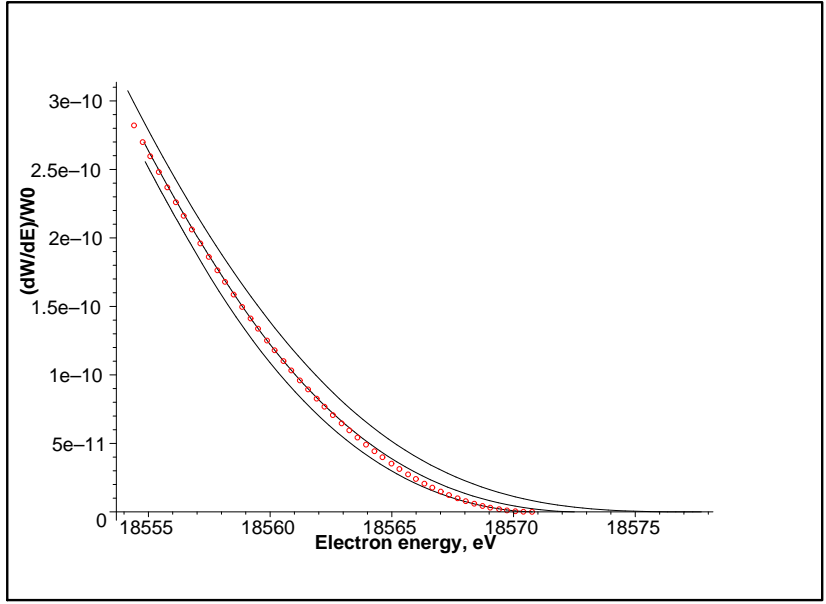}
\caption{Tritium beta spectrum near its end point: circles
correspond to $\xi=0,\;m_{\nu}=0;$ solid lines correspond to $\xi
=0.00005,$\, $m_{\nu} = 0;\;
5.1\,{\textmd{eV}};\;6.8\,{\textmd{eV}}.$}\label{fermi03}
\end{center}
\end{figure}

The behavior of the spectrum is illustrated by plots in Fig. \ref
{fermi1} - \ref {fermi03}, where $W_{0}$ is the total probability
of beta decay without external field.

 When $\xi\ll 1,$
\begin{equation}\label{p8}
t_{max} \approx \varepsilon_{0} -\mu
+\xi(\varepsilon_{0}^{2}-1)^{1/2}.
\end{equation}
Eq. (\ref{p8}) allows us to evaluate the field strength, which is
required to observe the effect. If the shift of the spectrum end
point induced by neutrino mass is equal to the shift induced by
external field, one has
\begin{equation}\label{p9}
  E\lambda= \frac{2\pi}{\sqrt{\varepsilon^{2}_{0}-1}}\,m_{\nu}.
\end{equation}
\noindent Here $E$ is the radiation field strength (V/m),
$\lambda$ is the wavelength of radiation (m), and $m_{\nu}$ is the
neutrino mass (eV).
\begin{figure}[H]
\begin{center}
\includegraphics[width=0.67\textwidth]{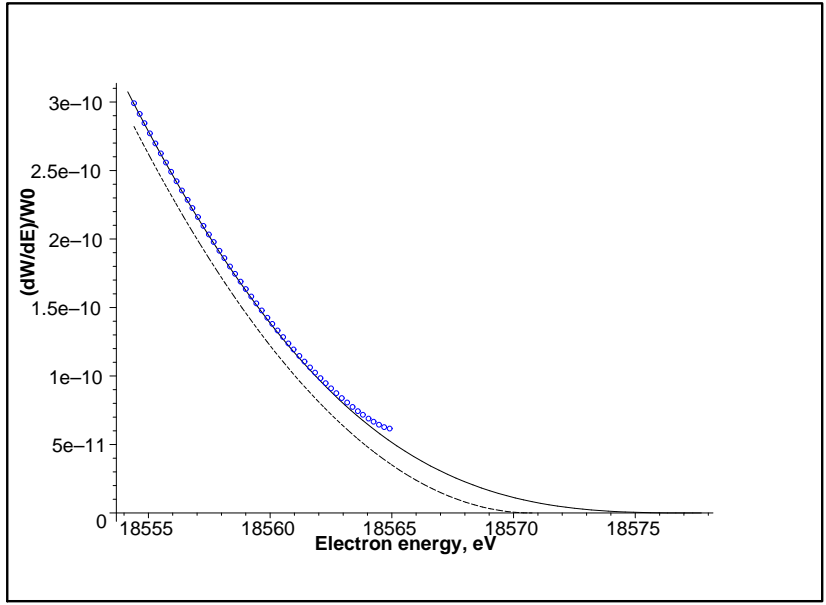}
\caption{Tritium beta spectrum near its end point: dashed line
corresponds to $\xi=0,$ $m_{\nu}=0;$\; solid line corresponds to
$\xi =0.00005, \;m_{\nu} = 0;$ circles correspond to $\xi=0,$ $
m_{eff}^{2}=-25\,{\textmd{eV}}^{2}.$}\label{fermi005}
\end{center}
\end{figure}\vspace{-8pt}
For tritium $\varepsilon_{0} \approx 1.03634,$ and the shift of
the spectrum end point induced by neutrino mass $\sim$1\,eV is
equal to the shift induced by microwave radiation with the
strength $\sim$10-100\,V/m.

For $\;\,t<t_{1}\,\;$ and $\,\;t>t_{1}\,\;$ analytical expressions
for beta spectrum are different. If $\mu =0,\; \xi \ll 1
,\;\varepsilon_{0} - t \ll 1,\;t>t_{1}$  the following
approximation exists (Fig. \ref{fermi005}):
\begin{equation}\label{p10}
\frac{d(W_{eff}/\Tilde{W})}{dt}=t\sqrt{t^{2}-1}
\left[2(\varepsilon_{0}-t)^{2}-(\varepsilon_{0}-t)
\sqrt{(\varepsilon_{0}-t)^{2}-\mu_{eff}^{2}}\right],
\end{equation}
\noindent where
\begin{equation}\label{p11}
\mu_{eff}^{2}=2\,\xi^{2}\left[\frac{\varepsilon_{0}}
{\sqrt{\varepsilon_{0}^{2}-1}}\ln\left(\varepsilon_{0}+
{\sqrt{\varepsilon_{0}^{2}-1}}\right)-1\right].
\end{equation}
\noindent This approximation was used for analyzing experimental
data \cite{WDB99,LAB99} (the so called "negative neutrino mass
squared"). Thus the electromagnetic radiation could be the reason
for experimentally observed anomaly in tritium beta spectrum.

\begin{figure}[H]
\begin{center}
\includegraphics[width=0.8\textwidth]{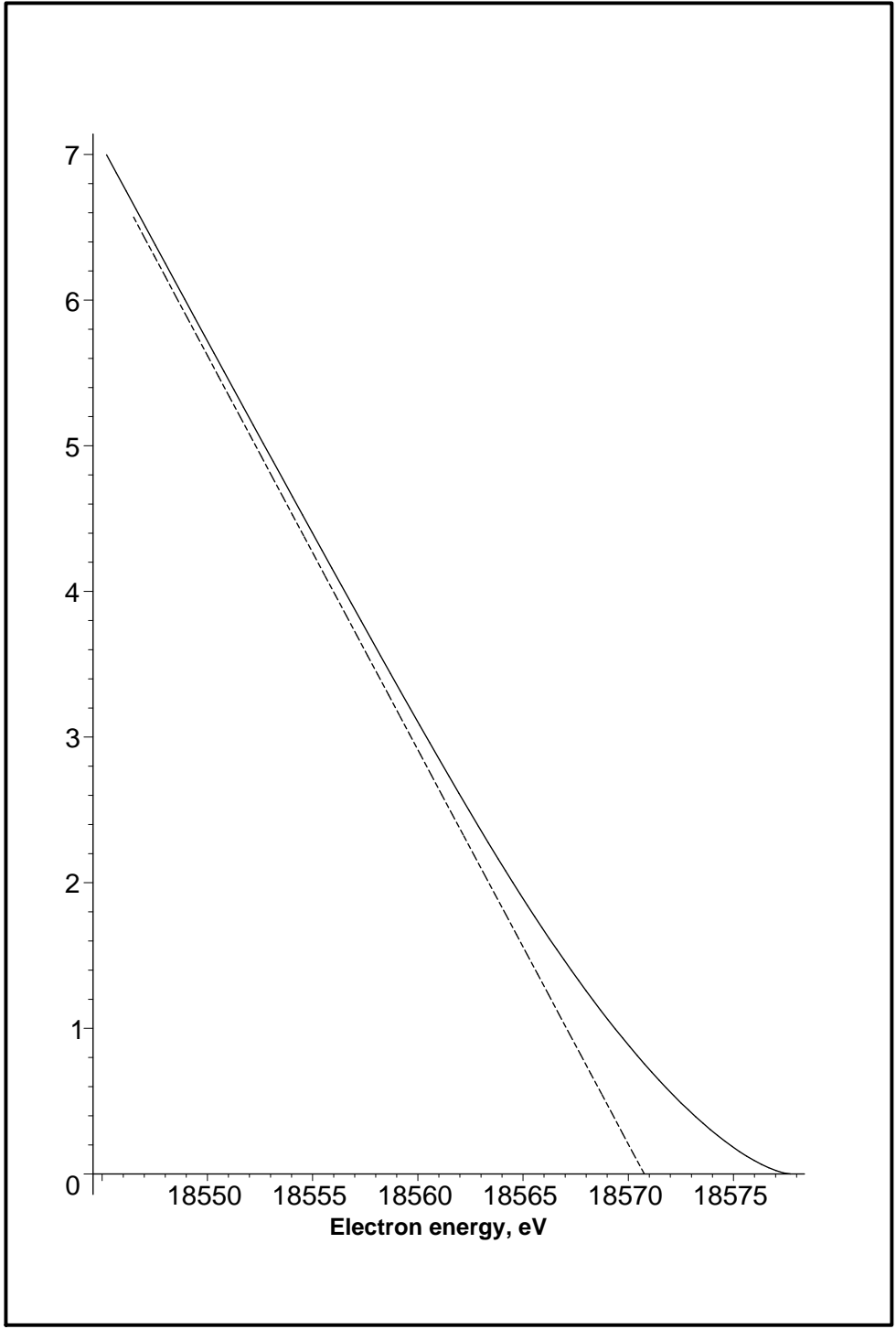}
\caption{Curi plot for tritium in arbitrary units: solid line
corresponds to $\;m_{\nu} =0,$ $\xi = 0.00005;\;$ dashed line
corresponds to $m_{\nu} =0,\;\xi = 0.$}\label{curi1}
\end{center}
\end{figure}

Using the Curi plot
\begin{equation}\label{p12}
  C\sim \sqrt{\frac{dW/dt}{t(t^{2}-1)^{1/2}}}\,,
\end{equation}
 we can make the representation of our results more obvious (Fig.
\ref{curi1}).

The numerical estimations demonstrate that the shift in the
tritium beta spectrum, which corresponds to neutrino mass
$\sim$1\,eV (the limiting accuracy of experiments
\cite{WDB99,LAB99}), can be compensated by a microwave radiation
field with the strength of the order of tens of V/m. In planned
experiment KATRIN \cite{K01,W02}, in which the measurement
accuracy is supposed to be of the order of 0.1\,eV, this can be
produced by fields with the strength of the order of units of V/m,
which is comparable with background values.

In the analysis of the experimental data, one should investigate
the role of possible external sources of radiation, as well as
radiation of beta electrons. In fact, since in chambers of
experimental setups there exist constant magnetic fields with the
magnitude of the order of units of Tesla, maximum of
$\beta$-electron electromagnetic radiation belongs to cm-range.

It should be noted that in a more detailed investigation it is
necessary to allow for effects related to the energy loss due to
transitions to excited states of molecular tritium (see
\cite{JM96}), but in our opinion this factor cannot change main
conclusions of the present work.


\begin{thebibliography}{99}

\bibitem{L7} Ternov I.M., Rodionov V.N., Lobanov A.E.,  and
Dorofeev~O.F. {\em Pisma v Zh. Eksp. Teor. Fiz.} 37 (1983) 288.
\bibitem{L10} Ternov I.M., Rodionov V.N., Dorofeev~O.F., Lobanov
A.E.,
and Pavlova O.S. {\em Yad. Fiz.} 39 (1984) 1125.
\bibitem{L18} Ternov I.M., Rodionov V.N., Zhulego V.G., Lobanov A.E.,
Pavlova O.S., and  Dorofeev~O.F. {\em J. Phys. G.} 12 (1986) 637.
\bibitem{WDB99} Weinheimer Ch., Degenddag B., Bleile A. et~al.
{\em Phys. Lett. B.} 460 (1999) 219.
\bibitem{LAB99}
 Lobashev~V. M., Aseev~V. N., Belesev~A. I. et~al.
{\em Phys. Lett. B.} 460 (1999) 227.
\bibitem{K01}
{KATRIN} {C}ollaboration, Osipowicz~A. et~al. {\em hep-ex/0109033}
(2001).
\bibitem{W02}
 Weinheimer Ch.
{\em hep-ex/0210050} (2002).
\bibitem{JM96}
Jonsell S. and Monkhorst~H. J. {\em Phys. Rev. Lett.} 76 (1996)
4476.

\end{thebibliography}
\end{document}